# Resonant driving of magnetization precession in a ferromagnetic layer by coherent monochromatic phonons

J. V. Jäger[1], A. V. Scherbakov[2], B. A. Glavin[3], A. S. Salasyuk[2], R. P. Campion[4], A. W. Rushforth[4], D. R. Yakovlev[1,2], A. V. Akimov[4], and M. Bayer[1,2]

[1]*Experimentelle Physik 2, Technische Universität Dortmund, D-44227 Dortmund, Germany*
[2]*Ioffe Physical-Technical Institute, Russian Academy of Science, 194021 St. Petersburg, Russia*
[3]*Department of Theoretical Physics, Lashkaryov Institute of Semiconductor Physics, 03028 Kyiv, Ukraine*
[4]*School of Physics and Astronomy, University of Nottingham, Nottingham NG7 2RD, UK*

ABSTRACT

We realize resonant driving of the magnetization precession by monochromatic phonons in a thin ferromagnetic layer embedded into a phononic Fabry-Perot resonator. A femtosecond laser pulse excites resonant phonon modes of the structure in the 10-40 GHz frequency range. By applying an external magnetic field, we tune the precession frequency relative to the frequency of the phonons localized in the cavity and observe the enormous increase in the amplitude of the magnetization precession when the frequencies of free magnetization precession and phonons localized in the cavity are equal.

The continual miniaturization of magnetic devices down to the nanometer scale has opened new horizons in data storage [1], computing [2,3], sensing [4,5], and medical technologies [6]. Progress in nanomagnetism is stimulated by emerging technologies, where methods to control magnetic excitations on the nanometer spatial and picosecond temporal scales include optical [7,8], electrical [8], and micromechanical [9] techniques. To realize ultrafast nanomagnetism on the technological level, new physical principles to efficiently induce and control magnetic excitations are required, and this remains a challenging task. A new basic approach to this problem would be to explore nanoscale magnetic resonance phenomena: resonant driving and monitoring of magnetic excitations, which is widely used nowadays in traditional magnetism for microscopy, medicine and spectroscopy. The typical frequencies, $f_M$, of magnetic resonances [e.g. the ferromagnetic resonance (FMR) in ferro- and ferrimagnetic materials], , are in the GHz and sub-THz frequency ranges. The traditional methods to scan magnetic excitations at these frequencies use microwaves, but due to the requirement of massive microwave resonators providing long wavelength radiation, they cannot provide high-speed control of magnetization locally on the nanoscale.

Among various emerging techniques in nanomagnetism, the application of stress to magnetostrictive ferromagnetic layers has been shown to be an effective, low power method for controlling magnetization: applying in plane stress in stationary experiments enables irreversible switching of the magnetization vector [10]; the injection of picosecond strain pulses induces free precession of the magnetization [11]; excitation of quantized elastic waves in a membrane enables driving of the magnetization at GHz phonon frequencies [12]; and surface acoustic waves can be used to control the magnetic dynamics in periodic ferromagnetic nanostructures [13,14]. In the present work, we examine the interaction of a high-frequency (10-40 GHz) magnetic resonance in a magnetostrictive ferromagnetic film with an elastic harmonic excitation in the form of a localized phonon mode, and demonstrate how this interaction becomes significantly stronger at the resonance conditions.

Our device consists of a ferromagnetic layer embedded into a phonon Fabry-Perot (FP) cavity. Such a cavity possesses quantized resonances for elastic waves (i.e. phonons) at frequencies $f_{Ri}$ ($I$ is the order of the phonon resonance). In the experiments, we excite these resonant modes optically by the methods of picosecond acoustics [15]. As it was shown earlier in experiments with picosecond strain pulses, coherent phonons induce free precession of the magnetization at the ferromagnetic resonance frequency $f_M$ [11,16]. By the application of an external magnetic field, **B**, we tune the frequency $f_M$ into resonance with the phonon mode, $f_M = f_{Ri}$, and monitor the precession of the magnetization in the ferromagnetic layer. We observe an enormous increase of the magnetization precession lifetime and the spectral density at $f_M$ when $B$ corresponds to matching the resonant conditions of the magnetization precession and the phonon cavity mode.

The studied structure is shown schematically in Fig. 1 (a). A 59 nm thick ferromagnetic layer of Galfenol (alloy of 81%Fe and 19%Ga [17,18]) capped by a 3-nm Al layer (to prevent oxidation) is deposited by magnetron sputtering onto acoustic Bragg mirrors. These mirrors are formed by two superlattices (SL1 and SL2) each consisting of 10 periods of GaAs/AlAs bilayers with thicknesses (in nm) 59/71 and 42/49 in SL1 and SL2, respectively. The SLs are grown by molecular beam epitaxy on a semi-insulating (001) GaAs substrate. The Galfenol layer plays the role of a FP cavity between two flat phonon mirrors: one mirror is the free surface and another is the corresponding SL1 or SL2 Bragg mirror. As SL2 is positioned below SL1, the cavity with the SL2 Bragg mirror therefore includes also all layers of SL1. The studied multilayer structure possesses a number of localized phonon modes with frequencies $f_{Ri}$, which fall into the stop-bands of the SLs [19-21]. Figures 1 (b) and 1 (c) show the dispersion curves calculated for longitudinal phonons in the studied SL1 and SL2, respectively. The lowest phonon stop-bands in SL1 and SL2 are centered at 20 GHz and 28 GHz, respectively. The horizontal bars in the zoomed fragments of Figs. 1 (b) and 1 (c) indicate the calculated frequencies of the lowest phonon FP modes for the SL1 cavity ($f_{R1} = 20.0$ GHz) and the SL2 cavity ($f_{R2} = 28.0$ GHz and $f_{R3} = 29.5$ GHz). The calculation of the FP phonon modes is described elsewhere [22].

Figure 1 (d) shows these phonon modes as well as magnetization precession frequency vs applied magnetic field for the studied sample. The squares show the measured dependence of the precession frequency in the studied Galfenol film on $B$, interpolated linearly by the solid line, which is in good agreement with previous experiments [18,23]. The horizontal dashed lines indicate the frequencies $f_{Ri}$ of the phonon modes in the SL1 and SL2 cavities. The intersections of the solid and dashed lines give the resonances $f_M(B) = f_{Ri}$, which occur at particular resonance magnetic fields $B=B_{Ri}$ marked in Fig. 1 (d) by the vertical arrows.

The experimental scheme is shown in Fig. 1 (a). The FP phonon cavity with the Galfenol layer was excited by optical pump pulses from an amplified Ti:Sapphire laser (duration 200 fs, wavelength of 800 nm, repetition rate 100 kHz) focused to a spot with diameter 100 μm at the sample surface. The maximum energy density of the pump pulse was ~ 10 mJ/cm$^2$. The probe pulses of lower density (~ 20 μJ/cm$^2$) split from the same laser and passed through an optical delay line were used to measure the transient magneto-optical Kerr rotation for monitoring the temporal evolution of the changes $\Delta M_z(t)$ of the z-projection of the macroscopic magnetization **M** of the Galfenol layer, i.e. of the magnetization component normal to the surface [24]. The sample was mounted in a helium cryostat with superconducting magnet. The experiments were performed at various temperatures up to room temperature and most of the results presented in the paper were obtained at $T$=170 K. The external magnetic field up to 700 mT was applied in the plane of the Galfenol layer, **B**||[100], which is close to the easy magnetization axis [10,18].

The temporal evolution of the detected signals $\Delta M_z(t)$ measured for three values of $B$ are shown in Fig. 2 (a). The signals possess oscillatory behavior with the period, amplitude and lifetime of the oscillations dependent on $B$. The most important result is the existence of a high-amplitude long-lived tail in the middle curve of Fig. 2 (a) taken at $B$=190 mT. This tail cannot be observed above the noise level at lower and higher fields (see the lower and upper curves, respectively).

To further inspect the data, we use a spectral domain presentation. Fast Fourier transform spectra (FFTs) of $\Delta M_z(t)$ for various $B$ are shown in Fig. 2 (b) in a frequency range 15 - 25 GHz. These spectra show a main spectral line centered at $f_1$=20.5 GHz which, taking into account the accuracy of determining the SL parameters, may be unambiguously associated with the localized phonon mode in the FP cavity at $f_{R1}$=20.0 GHz. The amplitude of this spectral line changes considerably with $B$: Fig. 2 (c) shows the $B$-dependence of this amplitude where we clearly see that the amplitude maximum takes place at $B$=180 mT. Figure 2 (b) also shows a broader spectral line with lower amplitude whose position shifts to higher frequencies with the increase of $B$. This line corresponds to the fast decaying free precession of the magnetization with frequency close to $f_M$, which $B$-dependence for our Galfenol film is shown in Fig. 1 (d). Comparing the coordinates of the first intersection in Fig. 1 (d) with the data in Figs. 2 (b) and 2 (c), we conclude that the maximum amplitude in the spectra is obtained at $B \approx B_{R1}$. This observation is the main result of the present work and demonstrates a resonance in the magnetization precession that is driven by coherent phonons with frequency $f_{R1}$.

We now discuss the observed increase of the spectral density at $f_M = f_{Ri}$ in more detail. The optical pump pulse absorbed in the Galfenol layer results in an instantaneous rise of the layer temperature, generating a broad spectrum of coherent phonons in the form of a picosecond strain pulse [15]. The major fraction of generated phonons leaves the Galfenol film with the sound velocity on a timescale of ~10 ps, but phonons with frequencies $f = f_{Ri}$ remain localized in the FP cavities for a longer time. The calculations for the cavity formed by SL1 at $f_{R1}$ gives a remarkably high value for the decay time of $\tau_{R1} \sim 10$ ns, which is three orders of magnitude longer than the escape time for non-resonant phonons from the cavity and two orders of magnitude longer than the lifetime of the free magnetization precession in this experimental geometry [18,23]. The localized phonons drive the magnetization precession at $f = f_{Ri}$ and this driving force will last during the leakage time $\tau_{R1}$. The amplitude of the precession amplitude will increase when the resonance condition $f_M = f_{R1}$ is fulfilled. This is clearly observed in our experiments at $B=B_{R1}=180$ mT for the signals measured in the temporal and spectral domains. The width $\Delta f$ of the resonant curve in Fig. 2 (c) is the same as the width of the spectral line for magnetization precession [23], which is in full agreement with the expectation for driving an object by a harmonic force. For non-resonant conditions, when $B$ differs remarkably from $B_{R1}$, the low intensity spectral line at $f \approx f_M$ corresponds to the quickly decaying free oscillations due to excitation of magnetic precession by phonons from the initially generated broad spectrum and instantaneous temperature rise [16,25].

Thus, we conclude that the spectral amplitude of the magnetic precession at the frequency of the driving force rapidly increases at the resonance condition $f_M = f_{R1}$ for $B \approx B_{R1}$. To show the general validity of this statement we perform further experiments for $B$-values at which $f_M$ falls into the region of the other two FP phonon modes of the SL2 cavity. In this case, the cavity, which comprises the Galfenol layer and SL1, has a length 1362 nm. The calculations for infinitely long SL2 predict two localized phonon states as shown in Fig. 1 (c). Figure 3 (a) shows the temporal evolution of $\Delta M_z(t)$ at $B=400$ mT. The signal lasts longer than 6 ns (the time interval available for the experimental measurements) and possesses pronounced beatings due to simultaneous excitation of several spectral components. The corresponding spectral lines are clearly seen in the frequency domain presentation in Fig. 3 (b). The frequencies of the peaks are independent of $B$, but their intensities vary strongly with $B$. The spectral lines with $f_2=28.4$ GHz and $f_3=30.0$ GHz show absolute maximal intensities in the magnetic field $B$ interval between 350 and 450 mT. The frequencies of these two spectral lines are in good agreement with those of the calculated phonon modes $f_{R2}=28.0$ GHz and $f_{R3}=29.5$ GHz and their maximum amplitudes are detected at the $B$– values of the intersections in Fig. 1 (d) demarking the resonance condition $f_M \approx f_{R2} \approx f_{R3}$. The origin of other lower amplitude spectral lines in Fig.2 (b) is the finite length of SL2. To confirm this we have calculated the spectrum of the lattice displacement, $G(\omega)$ near the surface assuming that light forming the excitation pulse is absorbed within a thickness $x_0=30$ nm of the Galfenol film:

$$G(\omega) \sim \left| \frac{1}{1+x_0^2 k_0^2} \left\{ i \frac{x_0 k_0 \left[1+R(\omega)e^{2ik_0 d}\right]}{\left[1-R(\omega)e^{2ik_0 d}\right]} + 1 \right\} \right| \quad (1)$$

where $\omega=2\pi f$, $k_0$ is the phonon wave vector in Galfenol, $d=59$ nm is the thickness of the Galfenol film, and $R(\omega)$ is the complex reflectivity of the acoustic wave incident from the Galfenol on SL1 and SL2 grown on the GaAs substrate. $R(\omega)$ is calculated according to Ref. [22]. The result of the calculations based on Eq. (1) is shown in Fig. 3 (c). Excellent agreement with the experiment is observed: there is a single line at $f=f_{R1}=20$ GHz and a number of spectral lines in the region of 25-32 GHz exactly as observed experimentally. In fact, the appearance of the additional spectral lines is due to the phonon reflection resonances in finite-thickness SL existing at frequencies beyond but close to the edges of the phonon stop-bands in a SL. Thus, we can conclude that resonant driving of the magnetization by coherent phonons also takes place for the SL2 cavity. The spectrum also consists visible nonzero background in the whole frequency range, which supplies the excitation of free precession at non-resonant magnetic fields.

Finally, we discuss the possible role of transverse (TA) phonons in the studied nanostructure. In the experiments, we do not see any resonant effects at $B-$values that correspond to the resonance of $f_M$ being equal to the frequency of localized TA phonons. This is not surprising because the optical pump excitation does not excite shear strain and correspondingly TA phonons in the present high symmetry geometry [26,27]. However, we could expect generation of LA and TA phonons from the magnetization precession, as has been observed in conventional microwave-driven magneto-acoustic experiments [28]. Then the magnetic and elastic resonance would form a coupled magneto-elastic excitation, resulting in renormalization of magnetic and phonon eigenstates [29]. The search for such excitations in ferromagnetic nanostructures is an extremely attractive field in nanomagnetism and nanomechanics.

In conclusion, we have demonstrated a new method to drive resonantly magnetization precession by sub-THz coherent phonons when the ferromagnetic layer is inserted into a phonon cavity. Resonant driving by coherent phonons can be used locally on the nanoscale and does not require external resonators in contrast to microwave excitation. Nowadays it is possible to generate coherent monochromatic phonons optically up to frequencies of ~ 1 THz [30-32], thus, opening appealing perspectives for resonant driving of magnetic excitations possessing resonances at higher frequencies, e.g. in antiferromagnetic nanostructures. Interaction of high-frequency phonons and magnons on the nanoscale can lead to the development of a new class of devices. For instance, the enhanced amplitude of the magnetization precession during resonant driving by coherent phonons shows the feasibility of precessional switching of the magnetization between metastable states [33].

We acknowledge the support of the work by the Deutsche Forschungsgemeinschaft and the Russian Foundation for Basic Research in the frame of the ICRC TRR 160 and by the grant BA 1549/14-1, the Government of Russia through the program P220 (14.B25.31.0025), the BMBF within the RELQUSA project (FKZ: 13N12462), and the Russian Academy of Science. AWR acknowledges support from a Career Acceleration Fellowship (EP/H003487/1), EPSRC, UK; AVA acknowledges the Alexander von Humboldt Foundation.

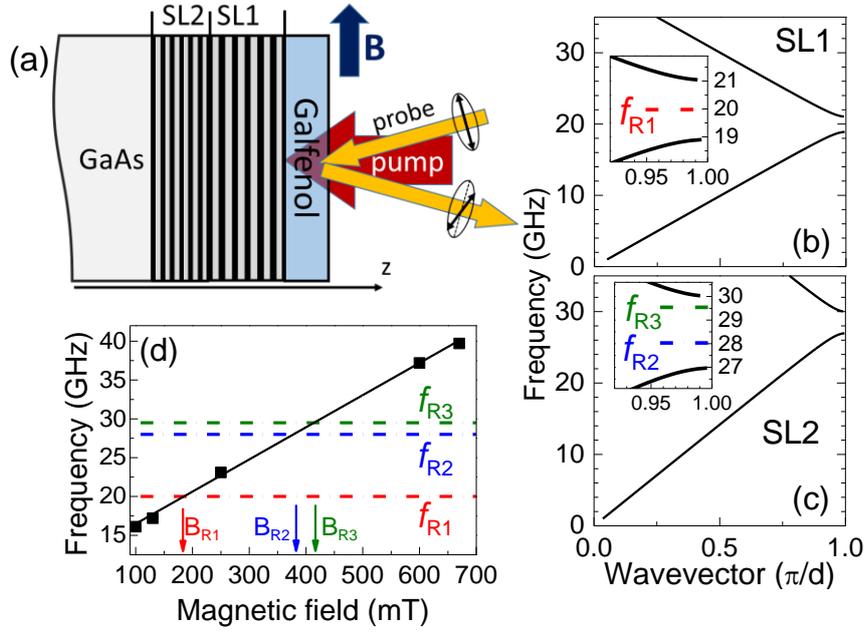

**Figure 1.** (a) The scheme of excitation and detection of magnetization precession in Galfenol layer grown on two phonon Bragg mirrors formed by GaAs/AlAs superlattices SL1 and SL2. (b), (c) Calculated phonon dispersion curves for two SL1 and SL2 Bragg mirrors, respectively. Horizontal bars in the zoomed insets indicate the frequency $f_{Ri}$ of the phonon localized modes formed by the free surface and the corresponding SL Bragg mirror. (d) The magnetic field dependence of the free precession frequency $f_M$ measured (symbols) and linear fit (solid line). Horizontal lines show the spectral positions of the cavity phonon modes $f_{Ri}$ and vertical arrows indicate the corresponding resonance fields $B_{Ri}$

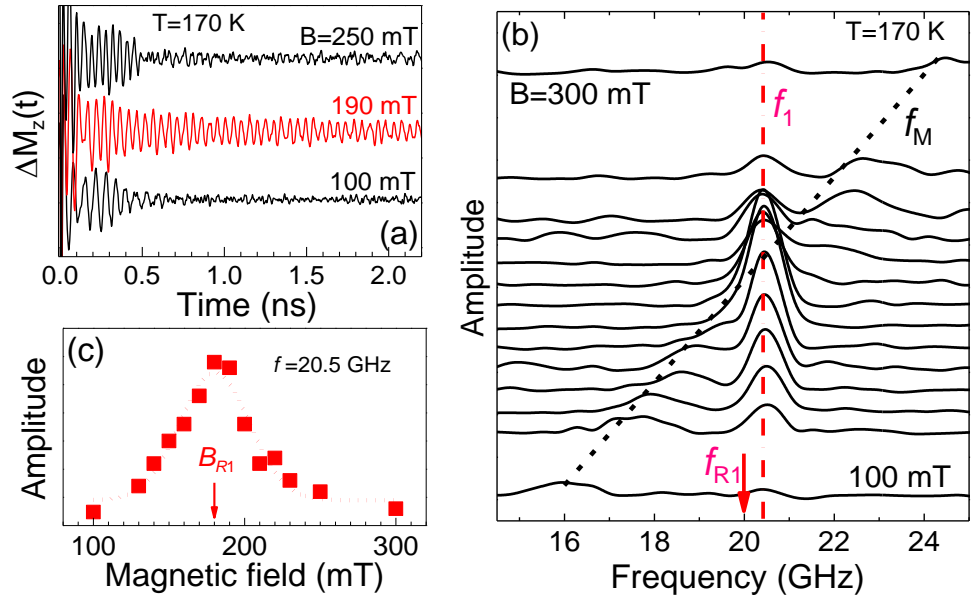

**Figure 2.** (a) Temporal evolution of the magnetization measured for three values of *B*. (b) The spectral density of the measured signals. Dashed vertical and oblique lines correspond to the maximum of the spectral line at $f_1$=20.5 GHz and estimated position of the free magnetic precession, $f_M$, respectively. The vertical arrow shows the calculated position of the lowest cavity mode, $f_{R1}$. (c) Measured (symbols) dependence of the spectral amplitudes on the magnetic field for the spectral line centered at $f_1$=20.5 GHz as indicated by dashed vertical line in (b); dotted line is corresponding Gaussian fit of the experimental data. Vertical arrow shows the value of the resonant magnetic field $B_{R1}$=180 mT.

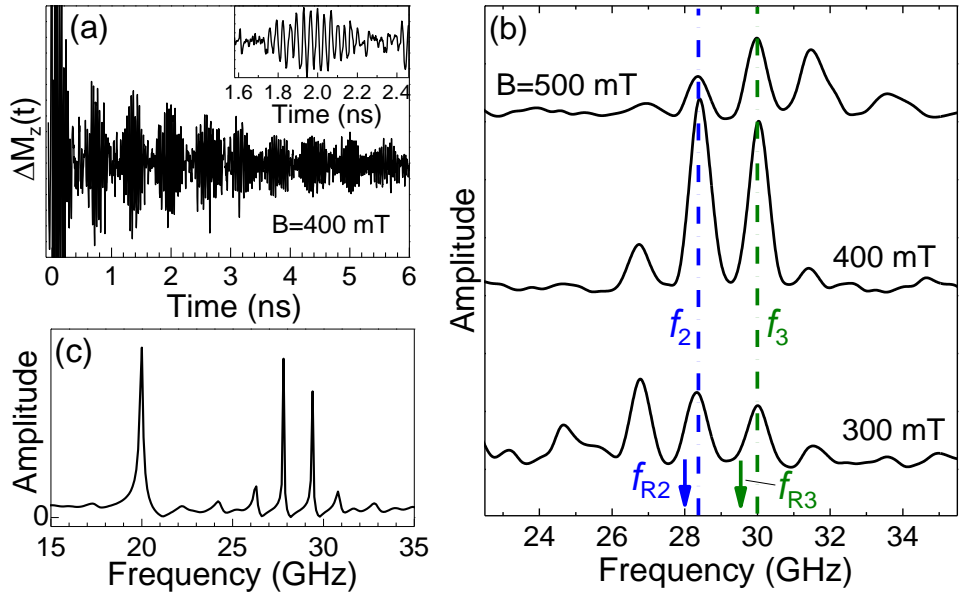

**Figure 3.** (a) Temporal evolution of the magnetization measured at $B$=400 mT, when the long living tail with beatings has the maximum amplitude. The inset shows the zoomed fragment of the measured signal. (b) Spectral density of the measured signals for three values of magnetic field. Vertical lines indicate the spectral positions $f_2$ and $f_3$ in the spectra of measured signals and vertical arrows show the calculated frequencies $f_{R2}$ and $f_{R3}$; (c) Calculated phonon spectrum for the studied structure for impulsive femtosecond optical excitation.